\newcommand{\ud}{\rm d}
\newcommand{\un}{~\mathrm}

\def\eg{ {\em e.g.} }

\def\EQ{\begin{equation}}
\def\EN{\end{equation}}
\def\EQA{\begin{eqnarray}}
\def\ENA{\end{eqnarray}}

\def\un{~\mathrm}
\def\bdes{\begin{description}}
\def\edes{\end{description}}

\documentclass[epj,final]{svjour}
\usepackage{graphicx}% Include figure files
\usepackage{dcolumn}% Align table columns on decimal point
\usepackage{bm}% bold math
\usepackage{amsmath}
\usepackage{hyperref}
\usepackage{psfrag}
\usepackage{color}

\begin{document}

\title{Euler-like modelling of dense granular flows: application to a rotating drum}

\author{D. Bonamy\inst{1} \and P.-H. Chavanis\inst{2} \and P.-P. Cortet\inst{1,3}
\and F. Daviaud\inst{3} \and B. Dubrulle\inst{3} \and M. Renouf\inst{4}}

\institute{CEA, IRAMIS, SPCSI, Grp. Complex Systems \& Fracture,
F-91191 Gif-sur-Yvette, France \and Laboratoire de Physique
Th\'eorique, CNRS UMR 5152, Universit\'e Paul Sabatier, 118 route
de Narbonne, F-31062 Toulouse, France \and CEA, IRAMIS, SPEC, CNRS
URA 2464, Groupe Instabilit\'es \& Turbulence, F-91191
Gif-sur-Yvette, France \and Equipe TMI, LaMCoS, CNRS UMR 5259,
INSA Lyon, 18-20 rue des sciences, F-69621 Villeurbanne, France}

\date{\today}

\abstract{General conservation equations are derived for 2D dense
granular flows from the Euler equation within the Boussinesq
approximation. In steady flows, the 2D fields of granular
temperature, vorticity and stream function are shown to be encoded
in two scalar functions only. We checked such
prediction on steady surface flows in a rotating drum simulated
through the Non-Smooth Contact Dynamics method. This result
is non trivial because granular flows are dissipative and
therefore not necessarily compatible with Euler equation.
Finally, we briefly discuss some possible ways to predict
theoretically these two functions using statistical mechanics.
\PACS{
    {47.57.Gc}{Granular flow}   \and
    {47.10.-g}{General theory in fluid dynamics} \and
      {83.80.Fg}{Granular solids}
     } % end of PACS codes
} %end of abstract

\maketitle

\section{Introduction}\label{intro}

The intrinsic dissipative nature of the interactions between the
constituent macroscopic particles sets granular media apart from
conventional solids, liquids and gases \cite{Jaeger96_rmp}.
Understanding the rheology of granular systems is thus rather
difficult. Depending on the flow velocity, three regimes are
usually distinguished: The {\em rapid flow} -- gaseous-like --
regime where grains interact through binary collisions, is
generally described within the framework of the kinetic
theory~\cite{Savage81_jfm,Jenkins83_jfm,Lun86_am}; The {\em slow
flow} -- solid-like -- regime, where grain inertia is negligible,
is most commonly described using the tools of soil mechanics and
plasticity theory \cite{nedderman92_book}. In between these two
regimes there exists a {\em dense flow} -- liquid-like -- regime
where grain inertia becomes important but contacts between grains
are still relevant. This last regime has been widely investigated
experimentally, numerically and theoretically (see
\cite{Gdrmidi04_epje} for a review) in various flow
configurations. Several constitutive laws have been derived by
accounting for non-local
effects~\cite{Mills99_epl,Andreotti01_pre,Jenkins02_pof,Bonamy03_epl,Rajchenbach03_prl},
by adapting kinetic
theory~\cite{Savage98_jfm,Bocquet02_pre,Mohan02_jfm}, by modelling
dense flows as partially fluidized flows~\cite{Aranson02_pre}, by
considering them as quasi-static flows where the mean motion
results from transient fractures modelled as self activated
process~\cite{Pouliquen96_pre,Debregeas00_epl,Pouliquen01_acs,Lemaitre02_prl}
or more recently by considering them as visco-plastic fluids
\cite{Iordanoff04_asmejt,Dacruz05_pre,Jop06_nature}. To our
knowledge, all these approaches fail to account for all the
features experimentally observed.

In some sense, similar difficulties are encountered in the
understanding and modelling of turbulent flows. In that case, the
challenge is to relate the Reynolds stresses, based on small scale
fluctuations, to large scales or time averaged quantities. A new
way to tackle this problem was recently suggested
\cite{Leprovost05_pre,Leprovost06_pre,Monchaux06_prl,Monchaux08_prl},
through the consideration of non-linear steady solutions of the
Euler equations, thereby disregarding any non-universal effects
induced by (large scale) forcing and (small scale) dissipation.
When applied to a turbulent von K{\'a}rm{\'a}n flow, this approach
leads to the characterization of the steady state velocity
fields through two scalar functions only, encoding all
information about the forcing and the dissipation. In the present
paper, this method is generalized to the case of inhomogeneous
dense granular flows. As a result, one obtains a characterization
of the steady state through two  scalar functions,
dependent on the forcing geometry and on the dissipation
processes, that relate the fields of granular temperature,
vorticity and stream function. In other words, the knowledge of
these two scalar functions is sufficient to encode the
two-dimensional (2D) hydrodynamical inhomogeneous fields.

The paper is organized as follows: In section \ref{theory}, the
structure of 2D steady granular flows is derived under some
specific assumptions. Hydrodynamics and state equations in
granular media are briefly discussed in section \ref{granhydr}.
Conservation equations are then rewritten assuming that volume
fraction is nearly constant within the flow (Boussinesq
approximation) in section \ref{bouss}, and then restricted to 2D
geometries in section \ref{2dcase}. In section
\ref{SectionStationary} the general shape of the stationary
solutions is given in the Euler approximation, assuming that, once
time-averaged, forcing and dissipation balance locally. In
particular, it is shown that these stationary states can be fully
characterized through the knowledge of two scalar functions $F$
and $G$. Section \ref{setup} confronts these predictions with
steady surface flows in rotating drum as obtained in Contact
Dynamics simulations reported in \cite{Renouf05_pof} that were
shown to reproduce the experimental features observed in
references \cite{Rajchenbach00_ap,Bonamy02_pof,Bonamy03_gm}. The
simulation scheme and the description of the simulated systems are
briefly recalled in section \ref{simmet}. Spatial distribution of
the averaged temperature, volume fraction, vorticity and stream
function fields are computed within the whole drum, at the grain
scale (Sec. \ref{results:I}). It appears that these
hydrodynamical fields can indeed be described through only two
scalar functions $F$ and $G$. This result is non trivial because
it tells us our granular dissipative flow is compatible with
non-dissipative Euler equation. The two characteristic functions $F$ and $G$ are
then determined from the numerical data (Sec. \ref{sec:closure}),
commented (Sec. \ref{discres}) and checked (Sec.
\ref{consistency}). In the last section of this paper (Sec.
\ref{discussion}) some possible ways to predict theoretically
these two functions are briefly discussed.

\section{Theoretical framework: Conservation equations within the Boussinesq approximation}\label{theory}

\subsection{Granular hydrodynamics}\label{granhydr}

It is commonly assumed that granular media can be described with
continuum models. In all the following, distances, time,
velocities and stresses are given in units of $d$, $d/g$,
$\sqrt{gd}$ and $\rho_0 g d$ respectively where $g$ refers to the
gravity constant, $d$ to the mean grain diameter, and $\rho_0$ to
the mass density of the grains. The mass, momentum and energy
conservation equations then lead to: \EQA
\partial_t\nu+\nabla \cdot \left (\nu {\bf v}\right) &=&0,\nonumber\\
\partial_t {\nu \bf v}
+\left({\bf v} \cdot \nabla\right) {\nu \bf
v}&=&-\nabla P+\nu {\bf g}+{\bf F}_{\textit{visc}}+{\bf F}_{\textit{forc}},\nonumber\\
\partial_t \nu T+\nabla\cdot \left(\nu T{\bf
v}\right) &=&-P\nabla \cdot{\bf v}+ E_{\textit{visc}} +
E_{\textit{forc}}. \label{conservation} \ENA

In these equations, $\nu({\bf r},t)$ is the field of volume
fraction; ${\bf v}({\bf r},t)$ is the coarse-grained velocity
field given by ${\bf v}({\bf r},t)=\langle {\bf c}_b(t)
\rangle_{b\in \Sigma({\bf r})}$ where ${\bf c}_b(t)$ refers to the
instantaneous velocity of the bead $b$ located at time $t$ within
the elementary volume $\Sigma({\bf r})$ located at position ${\bf r}$; $\bf g$ is
the gravitational acceleration; $T({\bf r},t)$ is the field of
granular temperature defined in term of the RMS part of the
velocity field, $T({\bf r},t)=\frac{1}{2}\langle ({\bf
c}_b(t)-{\bf v}({\bf r},t))^2 \rangle_{b\in \Sigma({\bf r})}$; ${\bf
F}_{\textit{forc}}({\bf r},t)$, $E_{\textit{forc}}({\bf r},t)$
denote the forcing (apart from gravity force) applying on this
elementary volume and ${\bf F}_{\textit{visc}}({\bf r},t)$,
$E_{\textit{visc}}({\bf r},t)$ stand for the dissipative processes
inside this elementary volume. This system has to be supplemented
by an equation of state $P=g(\nu)T$ and a rheology, \textit{i.e.}
some constitutive equations describing ${\bf F}_{\textit{forc}}$,
$E_{\textit{forc}}$, ${\bf F}_{\textit{visc}}$,
$E_{\textit{visc}}$.

Contrary to classical liquids, the density and temperature
dependence of transport coefficients play an important role in
determining the flow density. For dilute systems they are usually
obtained using kinetic theory of granular gases
\cite{Savage81_jfm,Jenkins83_jfm,Lun86_am} within the Enskog
approximation. For dense gases, there is no available systematic
theory allowing their description. They are therefore usually
prescribed using phenomenological models \cite{Bocquet02_pre} or
fitted using experimental
\cite{Dacruz05_pre,Jop06_nature,Jop05_jfm} or numerical
\cite{Speedy99_jcp} data. In particular, the equation of state can
be written in the high-density limit
\cite{Dacruz05_pre,Speedy99_jcp}: \EQ P\simeq
K\frac{\nu_\ast^2}{\nu_\ast-\nu} T, \label{equabizarre} \EN where
$K$ is a constant and $\nu_\ast$ the random close packing limit:
$\nu_\ast\simeq 0.82$ (resp. $\nu_\ast\simeq 0.64$) for 2D (resp.
for 3D) monodisperse packing. At $\nu=\nu_\ast$, this equation
therefore predicts a zero granular temperature, consistent with
the absence of motion. As for the dissipative terms and forcing,
the precise shape of the equation of state shall not be needed in the sequel.
This is a distinguished feature of our approach.

\subsection{The Boussinesq approximation}\label{bouss}

For simplicity, one focuses on situations where the volume
fraction is nearly constant close to the random close packing
limit $\nu\approx\nu_\ast$. In dense granular flows, this
approximation is generally satisfied within 10 percents
\cite{Gdrmidi04_epje}. Generalization to non constant volume
fraction is possible, but more involved. In that limit, the
classical Boussinesq approximation is implemented by neglecting
the fluctuation of volume fraction in the continuity equation so
that it becomes: \EQ \nabla \cdot {\bf v} \approx 0.
\label{continuity} \EN The other conservation equations may then
be simplified by defining a reference state with ${\bf v}=0$,
$T=0$, $P=P_\ast$, $\nu=\nu_\ast$, so that: \EQ \nabla
P_\ast=\nu_\ast {\bf g}, \label{hydrostatic} \EN \textit{i.e.} an
hydrostatic equilibrium in the vertical direction. Along with
non-zero velocity, we introduce temperature and volume fraction
deviations with respect to the reference state as: \EQ
\nu=\nu_\ast-\delta \nu; \quad T=\delta T; \quad P=P_\ast+\delta
P. \label{fluctu} \EN The momentum equation can then be written
as: \EQA
\partial_t {\bf v}+\left({\bf v} \cdot \nabla\right) {\bf v}&=&-\frac{1}{\nu}\nabla P+{\bf g}\nonumber +{\bf F}_{\textit{visc}}+{\bf F}_{\textit{forc}},\nonumber\\
&\approx &-\frac{1}{\nu_\ast}\nabla {\delta P}-\frac{\delta\nu}{\nu_\ast^2}\nabla P_\ast+{\bf g}-\frac{1}{\nu_\ast}\nabla P_\ast \nonumber\\
& & +{\bf F}_{\textit{visc}}+{\bf F}_{\textit{forc}},\nonumber\\
&=&-\frac{1}{\nu_\ast}\nabla {\delta
P}-\frac{\delta\nu}{\nu_\ast}{\bf g}+{\bf F}_{\textit{visc}}+{\bf
F}_{\textit{forc}}, \label{simplimomentum} \ENA where the
hydrostatic equilibrium has been used to simplify the last
equation. A similar treatment of the temperature equation leads
to: \EQ
\partial_t \delta T+\left({\bf v} \cdot
\nabla\right)\delta
T=E_{\textit{visc}}+E_{\textit{forc}}-\frac{P_\ast}{\nu_\ast}\nabla
\cdot {\bf v} \approx E_{\textit{visc}}+E_{\textit{forc}}.
\label{temperaturediff} \EN The system of resulting  equations can
be further transformed so that it involves only temperature
fluctuations by using equation (\ref{equabizarre}): \EQ
\frac{\delta\nu}{\nu_\ast}=\frac{\delta T}{T_{\textit{ref}}},
\label{thermo} \EN where the reference temperature field
$T_{\textit{ref}}({\bf r})$ is given by
$T_{\textit{ref}}=P_\ast/K\nu_\ast$ so that ${\bf g}=K \nabla
T_{\textit{ref}}$, and only the first order terms in $\delta
\nu/\nu_\ast$, $\delta T/T_{\textit{ref}}$ and $\delta P/P_\ast$
are kept. The system of equations of the weakly compressible
granular medium then takes the shape: \EQA
&&   \nabla \cdot {\bf v} = 0,\nonumber\\
&&\partial_t {\bf v}+\left({\bf v} \cdot \nabla\right) {\bf v} = -\frac{1}{\nu_\ast}\nabla {\delta P}-\frac{\delta
T}{T_{\textit{ref}}}{\bf g }+{\bf F}_{\textit{visc}}+{\bf F}_{\textit{forc}},\nonumber\\
&&\partial_t \delta T+\left({\bf v} \cdot \nabla\right) \delta T =
E_{\textit{visc}}+E_{\textit{forc}}. \label{conservation2} \ENA
Note that the system can also be formulated in a more classical
Boussinesq-like form by introducing the variable $\theta=\delta
T/T_{\textit{ref}}$ and noting that $T_{\textit{ref}}$ is not a
constant (it varies along the gravity direction), so that: \EQA
&& \nabla \cdot {\bf v} = 0,\nonumber\\
&& \partial_t {\bf v}
+\left({\bf v} \cdot \nabla\right) {\bf
v}=-\frac{1}{\nu_\ast}\nabla
 {\delta P}-\theta {\bf g}+{\bf F}_{\textit{visc}}+{\bf F}_{\textit{forc}},\nonumber\\
&& \partial_t \theta+\left({\bf v} \cdot \nabla\right)
\theta +\left({\bf v} \cdot \nabla\right) \log T_{\textit{ref}} = E_{\textit{visc}}+E_{\textit{forc}}.
\label{boussinesq}
\ENA
In the sequel, we shall however rather work with the formulation (\ref{conservation2}).

\subsection{2D case}\label{2dcase}

We now specialize our granular hydrodynamics to the case of 2D
medium, such as flow within a thin rotating drum of diameter $2R$,
rotated along the $y$ axis at a constant angular velocity $\Omega$
as investigated in section \ref{setup}. If the width of the drum
in the $y$ direction is thin with respect to the characteristic
length scale of $(x,z)$ motions, the velocity field can be assumed
two-dimensional ${\bf v}(x,z,t)$. In that case, the vorticity is
directed along the $y$ axis and the forcing is supplied by the
boundary conditions. One can recast equation (\ref{conservation2})
in cartesian coordinates $(x,z)$ as: \EQA
\partial_x v_x+\partial_z v_z &=& 0 \; ,\\
\partial_t v_x+v_x\partial_x v_x+v_z\partial_z
v_x&=&-\frac{1}{\nu_\ast}\partial_x \delta P -g_x
\frac{\delta T}{T_{\textit{ref}}}\nonumber\\
& & +F^x_{\textit{visc}}+F^x_{\textit{forc}}\nonumber \; ,\\
\partial_t v_z+v_x\partial_x v_z+v_z\partial_z v_z
&=&-\frac{1}{\nu_\ast}\partial_z \delta P-g_z
\frac{\delta T}{T_{\textit{ref}}}\nonumber\\
& & +F^z_{\textit{visc}}+F^z_{\textit{forc}}\; ,\nonumber\\
\partial_t \delta T+v_x\partial_x \delta T+v_z\partial_z
\delta T&=&E_{\textit{visc}}+E_{\textit{forc}} \; , \label{basic}
\ENA where $x$ and $z$ indices or superscripts denote the
components of the considered vector in a cartesian referential.
Thanks to incompressibility and the 2D nature of the flow, $v_x$
 and $v_z$ can be expressed in term of the stream function $\psi$ defined by:
\EQ v_x = \partial_z \psi \nonumber \qquad \text{and} \qquad v_z =
-\partial_x \psi \; . \label{stream} \EN Calling $q$ the
$y$-component of the vorticity, one gets: \EQ q=\partial_z
v_x-\partial_x v_z = \Delta\psi. \label{beau} \EN where
$\Delta=\partial^2_x +\partial^2_z$ is the Laplacian. Taking the
curl of the equation for velocity, equations (\ref{basic}) can be
recast as: \EQA \label{beautiful}
\partial_t \delta T+\lbrace \psi,\delta T\rbrace&=&E_{\textit{visc}}+E_{\textit{forc}} \; , \\ \nonumber
\partial_t q+\lbrace
\psi,q\rbrace &=& K\lbrace\log T_{\textit{ref}},\delta
T\rbrace+\nabla \times \left({\bf F}_{\textit{visc}}+{\bf
F}_{\textit{forc}}\right) \ENA where
$\lbrace\psi,\phi\rbrace=\partial_{z}\psi\partial_{x}\phi-\partial_{x}
\psi\partial_{z}\phi$ is the Jacobian. The relation between
gravity and $T_{\textit{ref}}$ was used to simplify the buoyancy
term. This formulation of the stratified Navier-Stokes equation
has to be supplemented by appropriate boundary conditions. Notice
that only two scalar fields are sufficient to describe the flows
under consideration: $\delta T$, the granular temperature and $q$,
the $y$-component of the vorticity.

\subsection{Steady state solutions}\label{SectionStationary}

Let us now consider steady regimes. At the global scale, the
dissipation generated by the interactions between grains should
balance exactly the external forcing applied by the drum on the
packing. From now, we {\em assume} that forcing and dissipation
equilibrate {\em locally} on average. This balance is all the more
likely since the considered elementary volume is large. In other
words, $\overline{{\bf F}_{\textit{visc}}(x,z,t)+{\bf
F}_{\textit{forc}}(x,z,t)}=0$ and
$\overline{E_{\textit{visc}}(x,z,t)+E_{\textit{forc}}(x,z,t)}=0$,
where the overlines denote averaging over time, and we focus on
the left-hand side of equations (\ref{beautiful}) to see the
implications on the  form taken by the  fields $\psi$, $q$ and
$\delta T$. The steady states then obey the averaged equations:
\EQA \overline{\lbrace \psi,\delta T\rbrace}&=&0 \; , \\ \nonumber
\overline{\lbrace \psi,q\rbrace} &=& K\lbrace \log
T_{\textit{ref}},\overline{\delta T}\rbrace. \; \label{average}
\ENA Neglecting correlations $\overline{\lbrace \psi,\delta
T\rbrace}\approx {\lbrace \overline{\psi},\overline{\delta
T}\rbrace}$, one gets: \EQA \lbrace \psi,\delta T\rbrace&=&0 \; ,
\\ \nonumber \lbrace \psi,q\rbrace &=& K\lbrace\log
T_{\textit{ref}},\delta T\rbrace \; , \label{tr2} \ENA where the
overlines over $q$, $T$ and $\psi$ are now omitted for sake of
simplicity. The first equation is satisfied if \EQA \delta
T=F(\psi), \label{tr3} \ENA where $F$ is an arbitrary function.
Using the general identity \EQA \lbrace f,h(g)\rbrace=h'(g)\lbrace
f,g\rbrace=\lbrace h'(g)f, g \rbrace, \label{tr4} \ENA where $f$,
$g$ and $h$ are arbitrary functions, the second equation becomes
\EQA \lbrace \psi,q+ F'(\psi)K \log T_{\textit{ref}}\rbrace=0.
\label{tr6} \ENA Therefore, the general stationary solution of
equations (\ref{beautiful}) is of the form \EQ \delta T = F(\psi)
\quad \text{and} \quad q+K F'(\psi) \log T_{\textit{ref}}=G(\psi),
\label{tr7} \EN where $F$ and $G$ are arbitrary functions.
Recalling the connection between $q$ and $\psi$, one can fully
characterize the stationary states through the two functions $F$
and $G$ as: \EQA
\delta T & = & F(\psi),\nonumber\\
\Delta\psi=q & = & -K\,F'(\psi)\log T_{\textit{ref}}+G(\psi).
\label{tr10} \ENA It should be emphasized that the functions $F$
and $G$ depend on the forcing and dissipation. Indeed, the
competition between these two effects are responsible for the
selection of the precise shape for $F$ and $G$. But once these
functions are known, one can solve the second equation of
(\ref{tr10}) to get $\psi$ as a function of $x$ and $z$, and then
derive from this expression the temperature and velocity profile.
To close the system of conservation equations, it is then
sufficient to give the expression for $F$ and $G$. There are
probably several ways to prescribe these functions. For example,
one could use a statistical mechanics approach in order to select
their ``most probable" form depending on macroscopic constraints and microscopic processes,
using methods of information theory (see \textit{e.g.}
\cite{Leprovost05_pre,Leprovost06_pre,Chavanis97_prl,Chavanis02_pre}
for illustrations in turbulence). One could also follow the
procedure used in  rheology studies, and try to define these
functions through ``minimal" experimental or numerical
measurements performed on the considered system.

\section{Application to simulated steady surface flows}\label{setup}

The formalism described in the previous section is now applied to
the inhomogeneous steady surface flows observed in rotating drums.

\subsection{Simulation methodology}\label{simmet}

The simulations have been performed using Non-Smooth Contact
Dynamics approach \cite{Moreau88_proc,Jean99_cmame}. The
algorithms benefit from parallel versions
\cite{Renouf04_cmame,Renouf04_jcam} which show their efficiency in
the simulation of large systems. The scheme has been described in
detail elsewhere \cite{Renouf05_pof} and is briefly recalled
below: An immobile drum of diameter $D_0=45\un{cm}$ is half-filled
with 7183 rigid disks of density $\rho_0=2.7\un{g.cm}^{-3}$ and
diameter uniformly distributed between $3$ and $3.6\un{mm}$. The
weak polydispersity introduced in the packing prevents 2D ordering
effects. The normal restitution coefficient between two disks
(resp. between disks and drum) is set to 0.46 (resp. 0.46) and the
friction coefficient to 0.4 (resp. 0.95). Once the packing is
stabilized, a constant rotation speed ranging from $2$ to
$15\un{rpm}$ is imposed to the drum. After one round, a steady
continuous surface flow is reached. One starts then to capture 400
snapshots equally distributed over one rotation of the drum.

\begin{figure}[]
\centerline{\includegraphics[width=0.9\columnwidth]{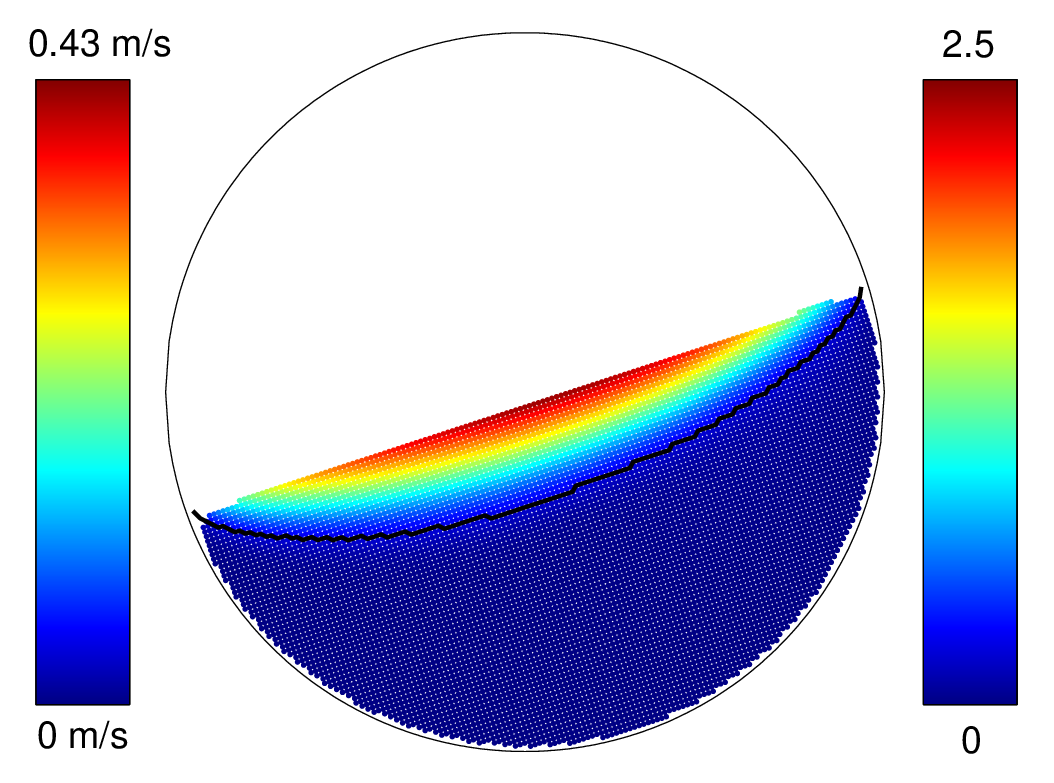}}
\caption{$x$-component of the time and ensemble averaged velocity
field in the simulated 2D rotating drum for a $\Omega=6\un{rpm}$
rotation speed. The black line shows the interface between the
flowing layer and the ``static" packing. Velocities are expressed
in m.s$^{-1}$ (left-hand colorbar) or non-dimensionalized by
$\sqrt{gd}$ (right-hand colorbar) where $g$ refers to the gravity
constant and $d$ to the mean diameter of the beads (see text for
details).} \label{snapshot2D}
\end{figure}

For each bead of each of the 400 frames within a given numerical
experiment, one records the position ${\bf r}$ of its center of
mass and its ``instantaneous" velocity ${\bf c}$ averaged over the
time step $\delta t = 6\times 10^{-3}\un{s}$ of the simulation.
For each rotation velocity, we have performed 20 experiments
starting from different initial packing of the beads. The
reference frame $\Re$ is defined as the frame rotating with the
drum that co{\"i}ncides with the reference frame $\Re_0=({\bf
e}_x,{\bf e}_z)$ fixed in the laboratory so that ${\bf e}_x$
(resp. ${\bf e}_z$) is parallel (resp. perpendicular) to the free
surface (Fig. \ref{snapshot2D}). The drum is then divided into
elementary square cells $\Sigma(x,z)$ of size set equal to the
mean bead diameter.

The average value of a field $a(x,z,t)$ at a
position $(x,z)$ is computed as a mixture of time and ensemble
average. Indeed, we performed averages of a quantity defined at
the grain scale over {\em all} the beads in {\em all} the 400
frames of {\em all} the 20 experiments whose center of mass is
within the cell located at $(x,z)$. Figure \ref{snapshot2D} shows
the spatial distribution of the $x$-component of the time-averaged
velocity field ${\bf v}(x,z)$ as obtained within this procedure.
The flowing layer and the static phase are then defined as the
point where $v_x$ is above and below a threshold value arbitrary
chosen to $0.2$. Let us note that all the results presented below
do not depend on this threshold value. The interface between the
two phases as defined within this procedure is represented as a
black line in figure \ref{snapshot2D}.

\subsection{Spatial distribution of the relevant continuous fields within the drum}\label{results:I}

\begin{figure*}
\psfrag{a}[c][][1.2]{(a)}
\psfrag{b}[c][][1.2]{(b)}
\psfrag{c}[c][][1.2]{(c)}
\psfrag{d}[c][][1.2]{(d)}
\psfrag{e}[c][][1.2]{(e)}
\psfrag{f}[c][][1.2]{(f)}
\centerline{\includegraphics[height=0.60\columnwidth]{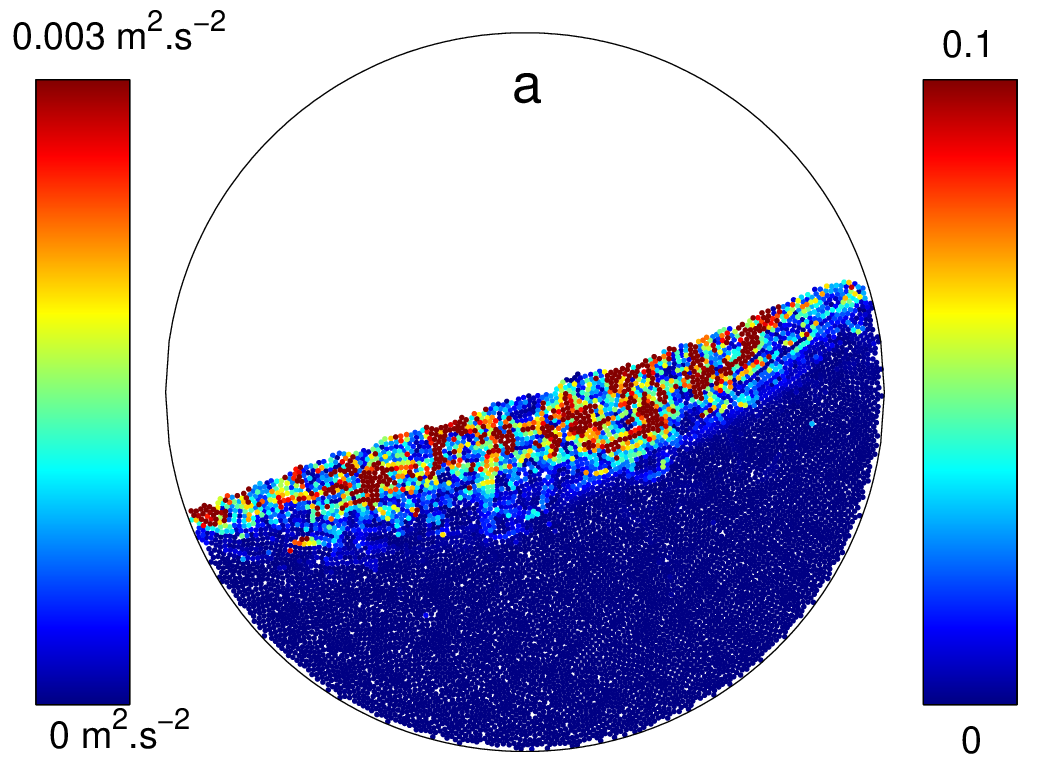}
\includegraphics[height=0.60\columnwidth]{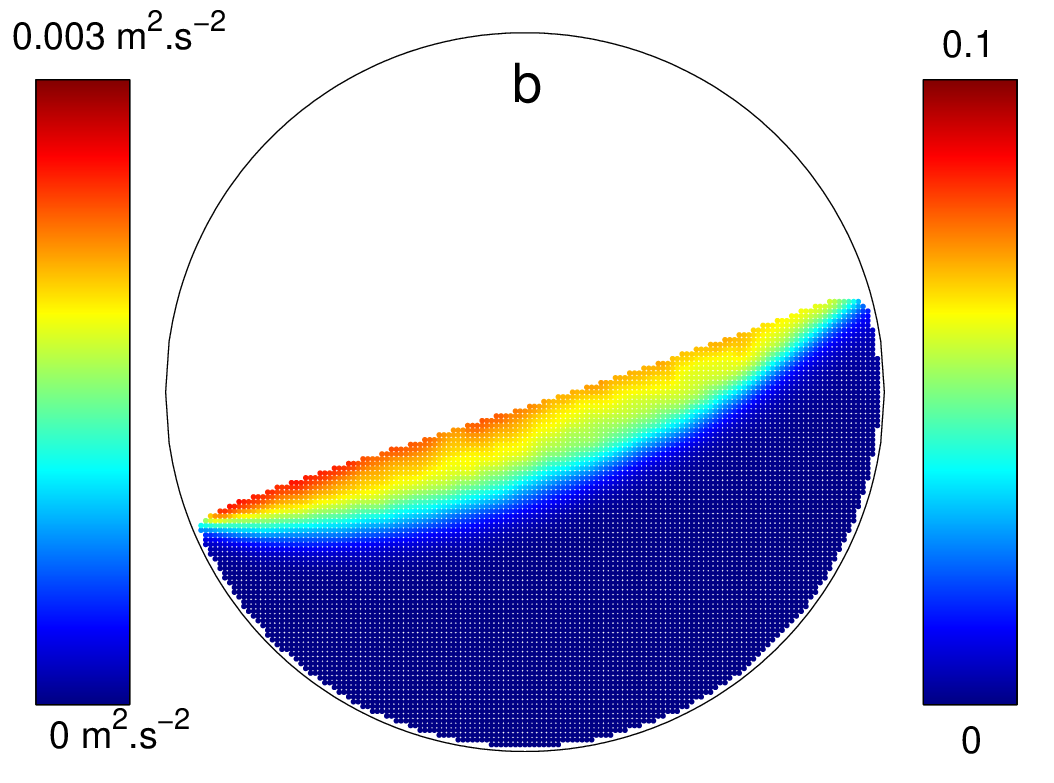}}
\centerline{\hspace{0.95cm}\includegraphics[height=0.60\columnwidth]{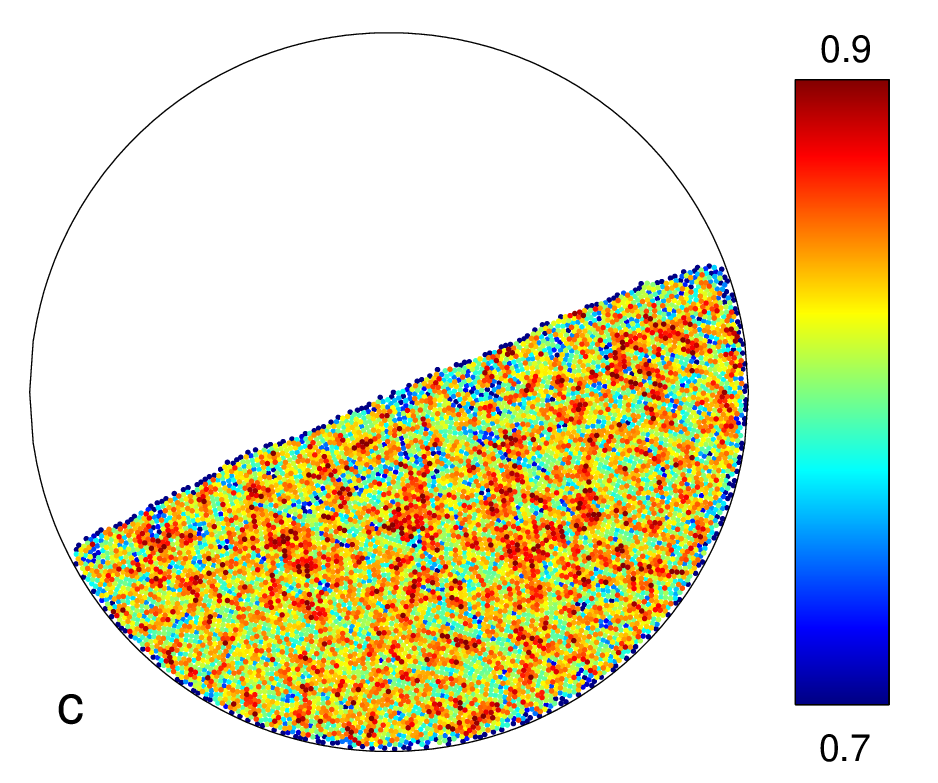}
\hspace{0.8cm}\includegraphics[height=0.60\columnwidth]{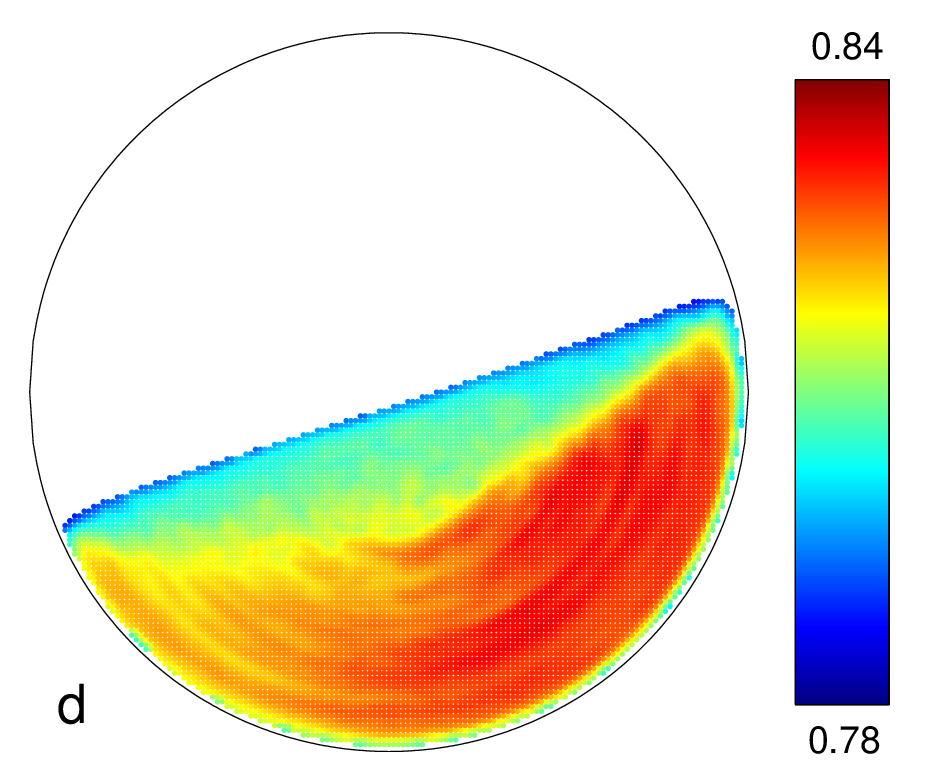}}
\centerline{\includegraphics[height=0.60\columnwidth]{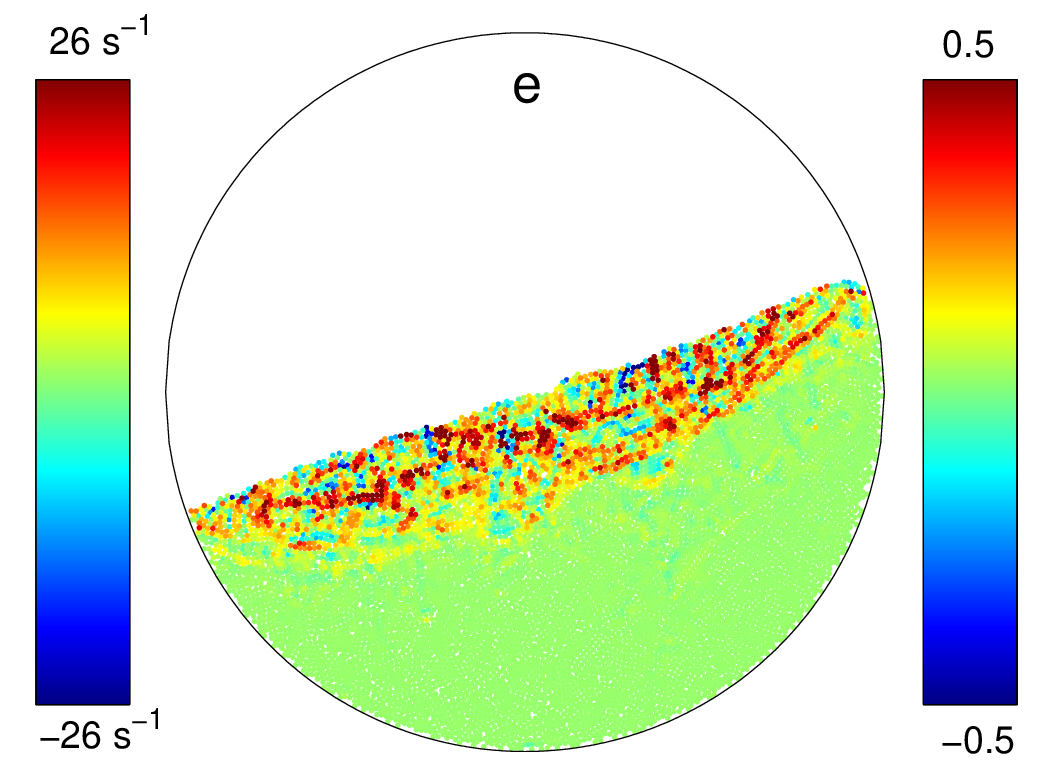}
\includegraphics[height=0.60\columnwidth]{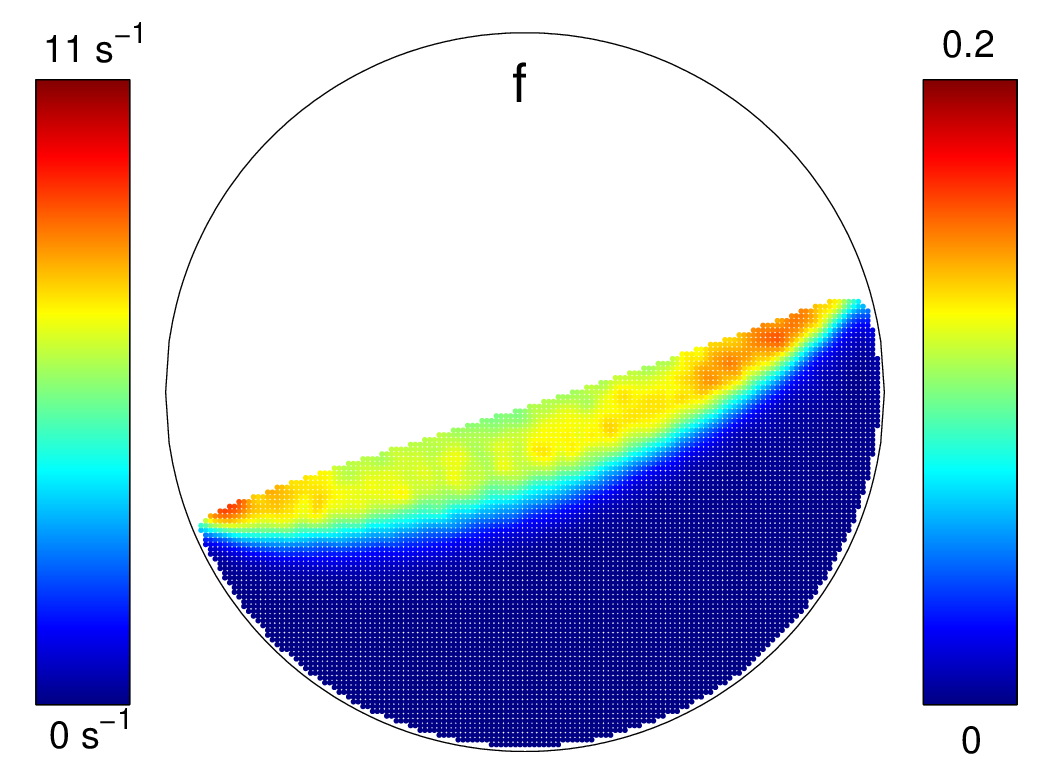}}
\caption{Spatial distribution of various continuous fields
measured within the drum for experiments with a $\Omega=6\un{rpm}$
rotation speed. Left: Typical snapshot of the instantaneous
spatial distribution of granular temperature (a), volume fraction
(c) and vorticity (e) within the rotating drum. Right: Time and
ensemble averaged field of temperature (b), volume fraction (d)
and vorticity (f). The average was taken over the 400 snapshots of
each of the 20 experiments for $\Omega=6\un{rpm}$. Temperatures
are expressed in m$^2$.s$^{-2}$ (left-hand colorbar) or
non-dimensionalized by $gd$ (right-hand colorbar). Vorticities are
expressed in s$^{-1}$ (left-hand colorbar) or non-dimensionalized
by $\sqrt{g/d}$ (right-hand colorbar) where $g$ refers to the
gravity constant and $d$ to the mean diameter of the beads (see
text for details).} \label{fields}
\end{figure*}

Let us first determine the granular temperature field within the
drum. Calling ${\bf c}_i(t)$ the instantaneous velocity of a bead
$i$ at a given time $t$, the fluctuating part of the velocity
$\delta {\bf c}_i(t)$ is defined as $\delta {\bf c}_i(t)={\bf
c}_i(t)-{\bf v}(x,z)$ where ${\bf v}(x,z)$ is the mean velocity
value on the cell $\Sigma(x,z)$ that contains the bead $i$. One
can then associate a granular temperature
$T_i(t)=\frac{1}{2}\delta c^2_i(t)$ to the considered bead. Figure
\ref{fields}a shows a typical snapshot of the instantaneous
temperature distribution within the drum as obtained using this
procedure. Two phases can be clearly distinguished. Within the
static phase, the temperature is very close to zero. Within the
flowing layer, the spatial distribution of instantaneous local
temperature shows large fluctuations, with hot and cold spots
gathered in transient clusters of various sizes. This structure of
hot and cold aggregates probably has its origin in the existence
of ``jammed'' aggregates embedded in the flow, as evidenced in
rotating drum experiments \cite{Bonamy02_prl}. Since we are
primarily interested in steady averaged fields in relation with
the theoretical framework developed in section \ref{theory}, we
focus on the spatial distribution of the temperature after
averaging over the 400 snapshots of each of the 20 experiments
performed for a given rotation velocity. The corresponding -- time
and ensemble -- averaged temperature field is represented in
figure \ref{fields}b.

Vorono{\"i} tessellation is then used to associate an
instantaneous elementary volume as defined in Continuum Mechanics
to each bead $i$ on each snapshot (see \eg \cite{Renouf05_pof} for
related discussion). Calling $A_i$ the area of the Vorono{\"i}
polyhedra enclosing the grain $i$, the instantaneous volume
fraction $\nu_i$ is defined as $\nu_i=\pi d_i^2/4A_i$ where $d_i$
denotes the diameter of bead $i$. Typical snapshot of the
resulting instantaneous map of volume fraction is presented in
figure \ref{fields}c. Apart from a very narrow region -- about one
bead diameter wide -- at the free surface and along the drum
boundary, the volume fraction appears almost constant, around
0.825, with apparent random fluctuations with standard deviation
around 0.04. However, the -- time and ensemble -- averaged field
of volume fraction presented in figure \ref{fields}d reveals that
$\nu(x,z)$ decreases slightly within the flowing layer, as
expected since dilatancy effects should accompany granular
deformation \cite{Reynolds85_pms}.

To compute the instantaneous vorticity $\omega_i$ associated to
each bead $i$ of each snapshot, the following procedure is
adopted: (i) The Vorono\"i polygon associated with the bead $i$ is
dilated homothetically by a factor two, so that each edge goes
through one of the neighboring beads'center; (ii) the circulation
$\Gamma_i(t)=\sum_{j}{\bf c}_j(t).{\bf s}_j(t)$ is calculated
around the resulting polygon -- each point of a given segment
${\bf s}_j$ is assumed to have a constant velocity ${\bf c}_j(t)$
equal to the one of the embedded bead; (iii) the instantaneous
vorticity $\omega_i(t)$ is then defined as
$q_i(t)=\Gamma_i(t)/A_i(t)$ where $A_i(t)$ refers to the area of
the initial Vorono\"i polygon.

A typical snapshot of the instantaneous vorticity distribution
within the drum as obtained using this procedure is presented in
figure \ref{fields}e. This distribution is complex. It exhibits
large fluctuations that self-organize into transient network of 1D
chains. The characterization of this transient structure is
postponed to future work. Figure \ref{fields}f presents the --
time and ensemble -- averaged vorticity field in the drum for
$\Omega=6\un{rpm}$.

The last continuous field of interest in relation with the
theoretical framework presented in section \ref{theory} is the
stream function $\psi(x,z)$. Its value is set to $\psi=0$ at the
drum boundary. The value $\psi(x,z)$ is then defined as the flow
rate going through a line connecting the point M at position ${\bf
r}(x,z)$ to any point at the drum boundary like \eg point M$_0$ at
position ${\bf r}_0(x,-\sqrt{D_0^2/4-x^2})$:

\begin{equation}
\psi(x,z)=\int_{-\sqrt{D_0^2/4-x^2}}^z v_x(x,u) \ud u
\end{equation}

\begin{figure}
\centerline{\includegraphics[width=0.9\columnwidth]{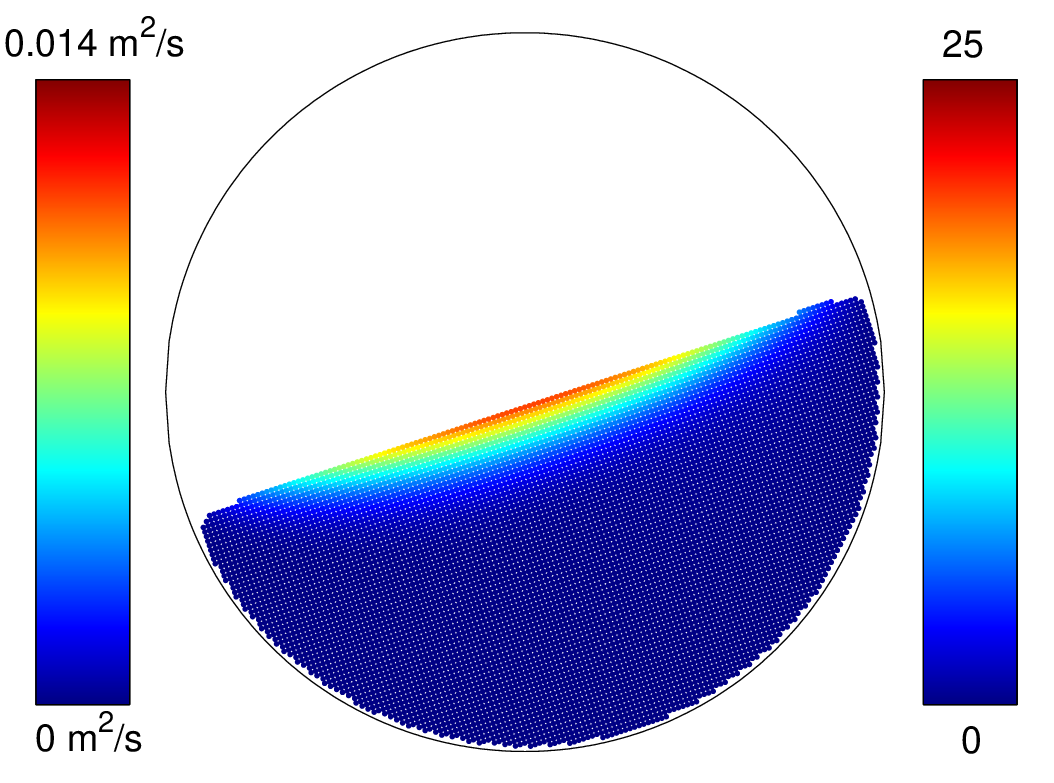}}
\caption{Spatial distribution of the time and ensemble averaged
stream function $\psi$ in the drum for $\Omega=6\un{rpm}$. The
stream function is expressed in m$^2$.s$^{-1}$ (left-hand
colorbar) or non-dimensionalized by $d\sqrt{gd}$ (right-hand
colorbar) where $g$ refers to the gravity constant and $d$ to the
mean diameter of the beads (see text for details).} \label{psi2D}
\end{figure}

The resulting spatial distribution of the stream function is shown
in figure \ref{psi2D}.

\subsection{Determination of the two scalar functions within Boussinesq approximation}\label{sec:closure}

\begin{figure}
\psfrag{n}[c][][1.3]{$\nu(x,z)$}
\psfrag{t}[c][][1.2]{$-T(x,z)/z\cos\theta$}
\centerline{\includegraphics[width=0.9\columnwidth]{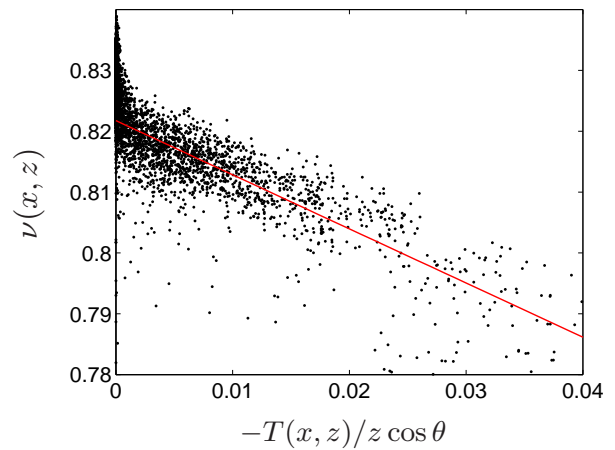}}
\caption{Time and ensemble averaged volume fraction $\nu(x,z)$ as
a function of the ratio $T(x,z)/z\cos\theta$ for
$\Omega=6\un{rpm}$. For this rotation speed, the mean slope of the
free surface was measured to be $\theta\simeq 19.7^\circ$
\cite{Renouf05_pof}. The straight line is a fit given by equation
(\ref{eq:nuvsT}) with $K \simeq 1.1$ and $\nu_*\simeq 0.822$. The
temperature is non-dimensionalized by $gd$ where $g$ refers to the
gravity constant and $d$ to the mean diameter of the beads (see
text for details).} \label{nuvsT}
\end{figure}

Let us first determine the value of the parameters $K$ and $\nu_*$
involved in the equation of state given by equation
(\ref{equabizarre}). This determination requires the pressure
field $P_*$ in the reference frame, when ${\bf v}={\bf 0}$. From
equation (\ref{hydrostatic}), one gets
$P_*(x,z)=-z\nu_*\cos\theta$ where $\theta$ is the slope of the
free surface. Equation (\ref{thermo}) can then be rewritten as:
\begin{equation}
\nu(x,z)=\nu_*+K\nu_*\frac{T(x,z)}{z\cos\theta}
\label{eq:nuvsT}
\end{equation}
The time and ensemble averaged local volume fraction $\nu(x,z)$ is
plotted as a function of the ratio $T(x,z)/z\cos\theta$ in figure
\ref{nuvsT}. The values of both $\nu_*$ and $K$ can then be
deduced. The volume fraction $\nu_*$ is found to be $\nu_*\simeq
0.824 \pm 0.003$ independently of the rotation velocity. The
parameter $K$ is found to be close to unity,  weakly dependent on
the rotating speed $\Omega$ \footnote{Strictly speaking, the
parameter $K$ is found to be significantly smaller for
$\Omega=2\un{rpm}$. However, for this particular rotating speed
the flowing layer is very thin. As a result, the variation range
of both $\nu(x,z)$ and $T(x,z)$ is very small and makes the fit
with equation (\ref{eq:nuvsT}) rather imprecise.} (see Tab.
\ref{tab1}). The reference temperature field
$T_{\textit{ref}}(x,z)=-z\cos\theta/K$ is then known.
\begin{table}[h]
\centering
\begin{tabular}{|c|cccccc|}
\hline
$\Omega$ & 2 rpm & 4 rpm & 5 rpm & 6 rpm & 10 rpm & 15 rpm \\
\hline
$K$ & 0.4 & 0.8 & 1.1 & 1.1 & 1 & 0.8\\
\hline
\end{tabular}
\caption{Variation of the parameter $K$ involved in the state
equation (\ref{equabizarre}) with respect to the rotating velocity
$\Omega$ of the drum. $K\simeq 1$ is found weakly dependent on the
rotating velocity $\Omega$.} \label{tab1}
\end{table}

\begin{figure}[h]
\psfrag{a}[c][][1.2]{(a)}
\psfrag{b}[c][][1.2]{(b)}
\psfrag{p}[c][][1.3]{$\psi(x,z)$}
\psfrag{P}[c][][1.3]{$\psi$}
\psfrag{f}[c][][1.3]{$F$}
\psfrag{t}[c][][1.3]{$T(x,z)$}
\centerline{\includegraphics[width=0.9\columnwidth]{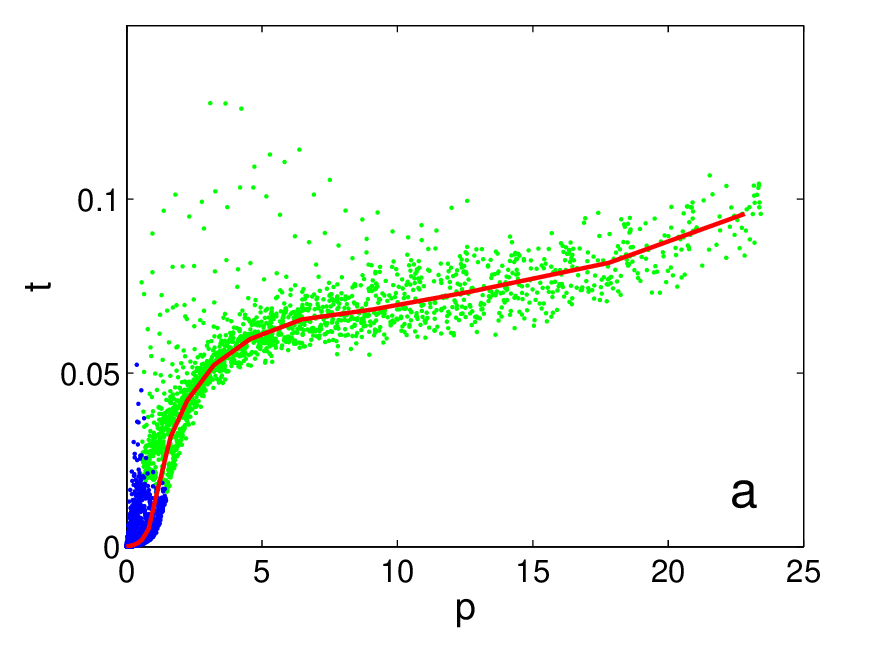}}
\centerline{\includegraphics[width=0.9\columnwidth]{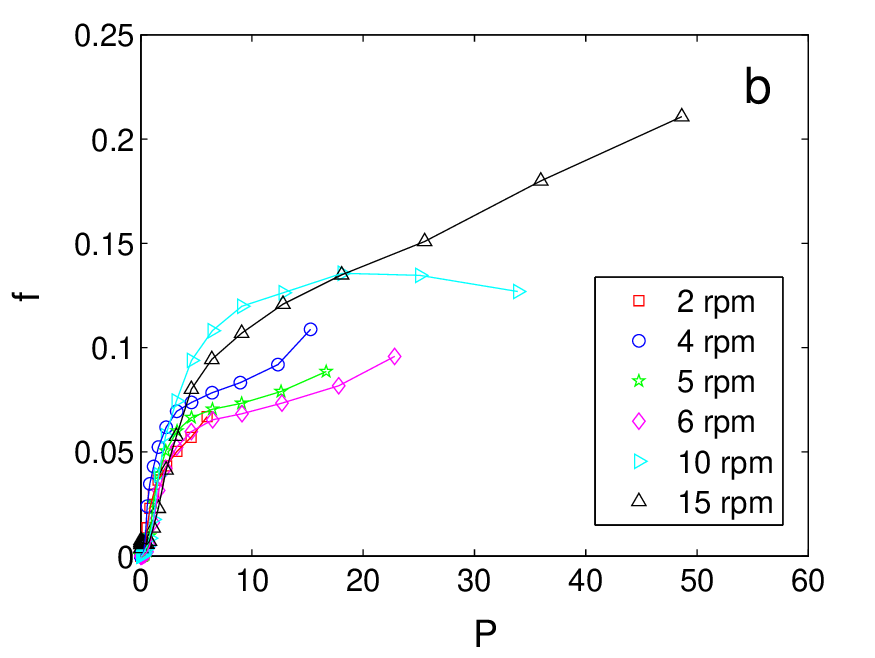}}
\caption{(a) Variation of the local temperature $T(x,z)$ as a
function of the local stream function $\psi(x,z)$ for
$\Omega=6\un{rpm}$. Each dot of the cloud corresponds to an
elementary square cell $\Sigma(x,z)$ of size equal to the mean
bead diameter. Green/light gray dots (resp. blue/strong gray dots)
correspond to points that belong to the flowing layer (resp. to
the static phase). The red line shows the function $\langle T
\rangle = F(\psi)$ where $\langle T \rangle$ is defined as the
average of the values $T$ for the cells $\Sigma(x,z)$ whose
$\psi(x,z)$ fall into logaritmically distributed bins. (b)
Variation of $F(\psi)$ as a function of $\Omega$. The temperature
and stream function are non-dimensionalized by $gd$ and
$d\sqrt{gd}$ respectively, where $g$ refers to the gravity
constant and $d$ to the mean diameter of the beads (see text for
details).} \label{Tvspsi}
\end{figure}

The knowledge of both the field $T(x,z)$ and $\psi(x,z)$ allows to
check the first equation in system (\ref{tr10}). Figure
\ref{Tvspsi}a shows $T(x,z)$ as a function of $\psi(x,z)$ for
$\Omega=6\un{rpm}$. The data points clearly gather along a single
function. It is worth to emphasize that such result would have
been trivial in unidirectional ``homogeneous'' flows such as
observed in plane shear or inclined plane geometry: in such flows,
all the continuum quantities depend on a single spatial coordinate
and are thus naturally related univocally by single functions. On
the contrary, the fact that the 2D fields $T(x,z)$ and $\psi(x,z)$
can be related by a {\em single} function in the inhomogeneous
multidirectional surface flow considered here, where the continuum
quantities depend on {\em both} spatial coordinates $x$ and $z$,
is highly non trivial and constitutes then a rather severe test
for the approach derived in section \ref{theory}. The function $F$
(red line in Fig. \ref{Tvspsi}a) is defined by averaging the
values $T$ falling into logarithmically distributed bins defined
along $\psi$. The functions $F$ obtained using this procedure for
the various rotating speeds $\Omega$ are represented in figure
\ref{Tvspsi}b.

\begin{figure}
\psfrag{a}[c][][1.2]{(a)}
\psfrag{b}[c][][1.2]{(b)}
\psfrag{p}[c][][1.3]{$\psi(x,z)$}
\psfrag{P}[c][][1.3]{$\psi$}
\psfrag{q}[c][][1]{$q(x,z)+KF'(\psi(x,z))\log T_{\textit{ref}}(x,z)$}
\psfrag{g}[c][][1.3]{$G$}
\centerline{\includegraphics[width=0.9\columnwidth]{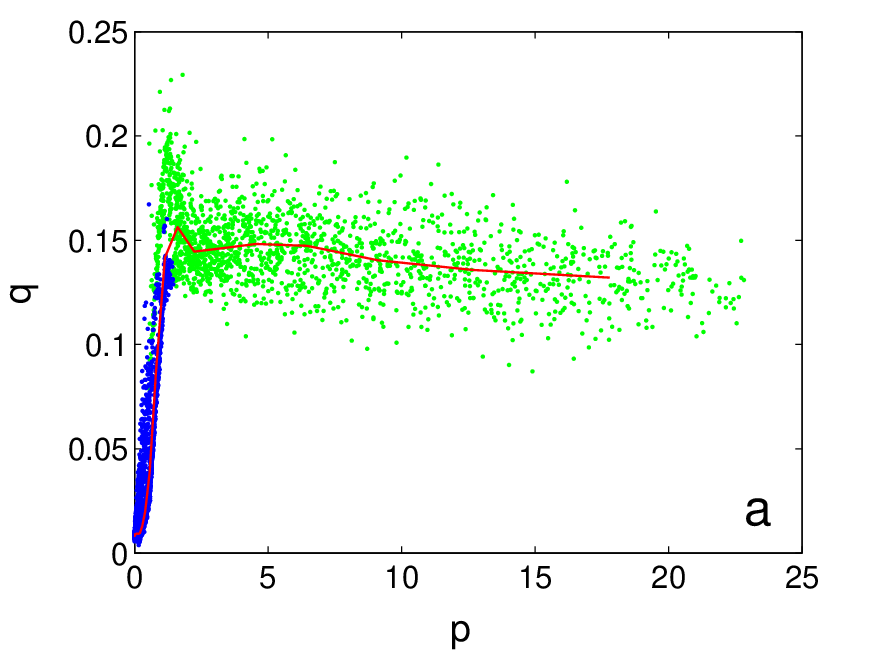}}
\centerline{\includegraphics[width=0.9\columnwidth]{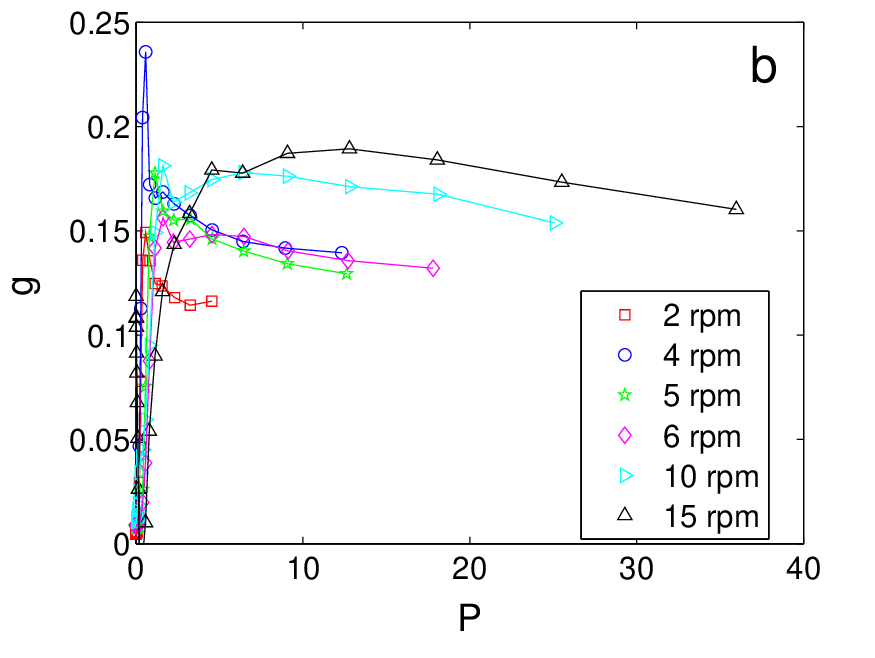}}
\caption{(a) Variation of the local field
$q(x,z)+KF'(\psi(x,z))\log T_{\textit{ref}}(x,z)$ as a function of
the local stream function $\psi(x,z)$ for $\Omega=6\un{rpm}$. Each
dot of the cloud corresponds to an elementary square cell
$\Sigma(x,z)$ of size equal to the mean bead diameter. Green/light
gray dots (resp. blue/strong gray dots) correspond to points that
belong to the flowing layer (resp. to the static phase). The red
line shows the function $\langle q+KF'(\psi)\log
T_{\textit{ref}}\rangle=G(\psi)$ where $\langle
\hspace{0.1cm}\rangle$ is defined as the average on the cells
$\Sigma(x,z)$ with values $\psi(x,z)$ which fall into
logaritmically distributed bins. (b) Variation of $G(\psi)$ as a
function of $\Omega$. The temperature, vorticity and stream
function are non-dimensionalized by $gd$, $\sqrt{g/d}$ and
$d\sqrt{gd}$ respectively, where $g$ refers to the gravity
constant and $d$ to the mean diameter of the beads (see text for
details).} \label{qvspsi}
\end{figure}

One can now determine the second closure relation $G(\psi)$. The
function $F(\psi)$ defined in the previous section (red line in
Fig. \ref{Tvspsi}a) is first derived numerically. The resulting
function is then applied at each point $(x,z)$ to the field
$\psi(x,z)$. Since the reference temperature
$T_{\textit{ref}}(x,z)=-z\cos\theta/K$ and the vorticity field
$q(x,z)$ are also known at each point, one can deduce the value of
the field $q(x,z)+KF'(\psi(x,z))\log T_{\textit{ref}}(x,z)$ at
each point, and plot it as a function of $\psi(x,z)$ (see Fig.
\ref{qvspsi}a). Again, the points clearly gather along a single
curve. The function $G$ (red line in figure \ref{qvspsi}a) is then
defined by averaging the values $q(x,z)+KF'(\psi(x,z))\log
T_{\textit{ref}}(x,z)$ falling into logarithmically distributed
bins defined along $\psi$. The functions $G$ obtained using this
procedure for the various rotating speeds $\Omega$ are represented
in figure \ref{qvspsi}b.

\subsection{Discussion of the results}\label{discres}

Our determination of the two scalar functions calls for some
comments. A first noticeable feature is that the function
extends smoothly, without any noticeable transition, from the
static to the flowing region. This is quite remarkable, since both
phases are characterized by different dynamical properties, and
since our hydrodynamic description presumably applies best within
the flowing region. The main difference between the two phases is
in the scattering of the data along the fit: It is larger in the
flowing region than in the static region. This may be traced to
correlated fluctuations that have been neglected in our approach
(see after Eq. (15)) and that are larger in the flowing region. It
would be interesting to see if a larger statistics leads to a
reduction of this scattering.

An interesting comparison can also be made with respect to a real
fluid system, where a similar approach can be used and where
dissipation is made through ordinary viscosity. In that case, it
has been shown in \cite{Monchaux06_prl} that the determination of
the scalar function is valid only in the bulk flow region.
Outside this region, the data scatters randomly, without forming
any specific shape. A possible explanation was that outside the
bulk, \textit{i.e.} closer to the boundaries and the flow forcing
devices, viscous and forcing processes become important and do not
balance {\em locally}  on average  as assumed here. The reason
why it works so well in the granular case, without any need of
selecting any flow region, may lie in the local character of the
dissipative processes that precludes any long-range correlation
between forcing and dissipation.

\subsection{Consistency check}\label{consistency}

As a consistency check, we can use the experimental curve $F$ and
$G$ to recompute the velocity and temperature fields and check
that they agree with profiles obtained in a rotating
drum. From figures \ref{Tvspsi} and \ref{qvspsi}, one sees that,
in that phase, $F$ is asymptotically linear $F\sim a\psi$, while
$G$ is approximately constant, $G\sim b$. Inserting these shapes
in equation (\ref{tr10}) leads to: \EQA
\delta T&=&a\psi,\nonumber\\
\Delta\psi&=&-aK\log (-z\cos\theta/K)+b \label{rheoaltgaz} \ENA
Integrating the second equation with respect to $z$, one finds:
\EQ \psi=\left( b - a K \log(\cos\theta/K)\right)\frac{z^2}{2}
+\frac{3}{2}aKz -\frac{aK}{2}z^2\log(-z) \label{integrgas} \EN so
that the temperature profile is quadratic, with logarithmic
correction and the velocity profile is linear, with logarithmic
correction. This is indeed the behavior observed in our rotating drum and, more generally,  in this type of
flow in the flowing phase
\cite{Gdrmidi04_epje,Rajchenbach03_prl,Renouf05_pof,Bonamy02_pof}.

In the static phase, $F$ appears quadratic in $\psi$, $F\sim c
\psi^2$ and $G$ is linear $G\sim d\psi$. Therefore, equation
(\ref{tr10}) becomes: \EQA
\delta T&=&c\psi^2,\nonumber\\
\Delta\psi&=&(d-2cK\log T_{\textit{ref}})\psi
\label{rheoaltgaz2}
\ENA
The solution for $\psi$ is in this case
\EQA
\psi&=&\psi_0 \exp(h(z)),\nonumber\\
h'^2(z)+h''(z)&=& d-2cK\log T_{\textit{ref}}>0, \label{integrstat}
\ENA so that both the velocity profile and the temperature
profiles are exponential, with algebraic corrections. This is
indeed the behavior observed in the static phase of our rotating drum or other similar type of
flows
\cite{Bonamy02_pof,Komatsu01_prl,CourrechdePont05_prl,Crassous}.

\section{Concluding discussion}\label{discussion}

In this paper, we investigate the steady states in 2D dense
granular flows within the Boussinesq-Euler approximation, assuming
local balance between time-averaged forcing and dissipation
exerted on an elementary volume.  We derived specific relations
between the continuum fields (temperature, vorticity and stream
function). In particular, we show that the fully 2D steady states
can be completely encoded in two scalar functions $F$ and $G$.
This prediction is then successfully checked onto the
stationary states of a dense inhomogeneous multidirectional
granular flow in a rotating drum. This means that stationary
states of the rotating drum can be described by a pure Euler
description, where neither the forcing, nor the dissipation are
explicitly taken into account. In the strict Euler equation framework, both
$F$ and $G$ would supposedly be determined by boundary and initial
conditions. However, in our approach, these conditions are only
effective and the functions $F$ and $G$ account implicitly for the
dissipation processes and the forcing geometry of the considered
forced dissipative flow. In this sense, the two scalar functions
$F$ and $G$ can be seen as fully encoding the 2D fields for
temperature and velocity in our apparatus. This represents a
reduction of the complexity of the description of the rotating
drum granular flows.

The main question in the present framework is therefore now to
understand and predict the shape of $F$ and $G$ as a function of
the forcing and dissipation. From an experimental or numerical
point of view, one may try and find empirical laws from variation
of the control parameters like rotation speed, size of the beads,
friction coefficient, etc. From a theoretical point of view, it
would be very interesting to be able to derive these functions
from a systematic theory. In a forthcoming paper, we explore a
strategy, based upon the statistical mechanics. This will lead to
a selection of the possible shapes of $F$ and $G$ based on
conservation laws and maximisation of an information entropy.
Moreover, this strategy leads to Gibbs distributions providing a
direct link between the function $F$ and $G$ and the fluctuations
of physical quantities. Therefore, from the knowledge of the mean
flow, one will be able to predict the velocity fluctuations. In
this respect, the present approach provides a useful insight for
dense granular flow in rotating drum and could be applied to other
granular flows. Finally, we stress that the present approach
relies heavily on the 2D character of the flow, that allows the
description of the flow non-linearities in terms of Jacobian. This
feature can be easily generalized to the case of 3D flows with
symmetries \cite{Leprovost06_pre}. Its extension to arbitrary 3D
geometry is currently the subject of a very active research.

\thanks{We gratefully acknowledge O. Dauchot for a critical reading of the
manuscript. Simulations are performed using LMGC90 software. This
work is supported by the CINE ({\em Centre d'Information National
et d'Enseignement}) under the project lmc2644. We are grateful to
S. Auma\^{i}tre, O. Dauchot, F. Leschenault and R. Monchaux for
many enlightening discussions.}


\begin{thebibliography}{10}

\bibitem{Jaeger96_rmp} H.M. Jaeger, S.R. Nagel, R.P. Behringer, Rev. Mod. Phys. \textbf{68}(4), 1259 (1996)

\bibitem{Savage81_jfm} S.B. Savage, D.J. Jeffrey, J. Fluid Mech. \textbf{110}, 255 (1981)

\bibitem{Jenkins83_jfm} J.T. Jenkins, S.B. Savage, J. Fluid Mech. \textbf{130}, 187 (1983)

\bibitem{Lun86_am} C.K.K. Lun, S.B. Savage, Acta Mech. \textbf{63}, 15 (1986)

\bibitem{nedderman92_book} R.M. Nedderman, \textit{Statics and Kinematics of Granular
Materials} (Cambridge University Press, Cambridge, 1992)

\bibitem{Gdrmidi04_epje} G.D.R. Midi, Eur. Phys. J. E \textbf{14}, 341 (2004)

\bibitem{Mills99_epl} P. Mills, D. Loggia, M. Texier, Europhys. Lett. \textbf{45}, 733 (1999)

\bibitem{Andreotti01_pre} B. Andreotti, S. Douady, Phys. Rev. E \textbf{63}, 031305 (2001)

\bibitem{Jenkins02_pof} J.T. Jenkins, D.M. Hanes, Phys. Fluids \textbf{14}, 1228 (2002)

\bibitem{Bonamy03_epl} D. Bonamy, P. Mills, Europhys. Lett. \textbf{63}, 42 (2003)

\bibitem{Rajchenbach03_prl} J. Rajchenbach, Phys. Rev. Lett \textbf{90}, 144302 (2003)

\bibitem{Savage98_jfm} S.B. Savage, J. Fluid Mech. \textbf{377}, 1 (1998)

\bibitem{Bocquet02_pre} L. Bocquet, W. Losert, D. Schalk, T.C. Lubensky, J.P.
Gollub, Phys. Rev. E \textbf{65}(1), 01307 (2002)

\bibitem{Mohan02_jfm} L.S. Mohan, K.K. Rao, P.R. Nott, J. Fluid Mech \textbf{457}, 377 (2002)

\bibitem{Aranson02_pre} I.S. Aranson, L.S. Tsimring, Phys. Rev. E \textbf{65} 061303 (2002)

\bibitem{Pouliquen96_pre} O. Pouliquen, R. Gutfraind, Phys. Rev. E \textbf{53}(1), 552 (1996)

\bibitem{Debregeas00_epl} G. Debregeas, C. Josserand, Europhys. Lett. \textbf{52}, 137 (2000)

\bibitem{Pouliquen01_acs} O. Pouliquen, Y. Forterre, S.L. Dizes, Adv. complex System \textbf{4}, 441 (2001)

\bibitem{Lemaitre02_prl} A. Lemaitre, Phys. Rev. Lett. \textbf{89}, 064303 (2002)

\bibitem{Iordanoff04_asmejt} I. Iordanoff, M.M. Khonsari, ASME J. Tribol. \textbf{14}, 341 (2004)

\bibitem{Dacruz05_pre} F. Da Cruz, S. Eman, M. Prochnow, J.-N. Roux, F. Chevoir, Phys. Rev. E \textbf{72}, 021309 (2005)

\bibitem{Jop06_nature} P. Jop, Y. Forterre, O. Pouliquen, Nature \textbf{441}, 727 (2006)

\bibitem{Leprovost05_pre} N. Leprovost, B. Dubrulle, P.-H. Chavanis, Phys. Rev. E \textbf{71}, 036311 (2005)

\bibitem{Leprovost06_pre} N. Leprovost, B. Dubrulle, P.-H. Chavanis, Phys. Rev. E \textbf{73}, 046308 (2006)

\bibitem{Monchaux06_prl} R. Monchaux, F. Ravelet, B. Dubrulle, A. Chiffaudel, F. Daviaud, Phys. Rev. Lett. \textbf{96}, 124502 (2006)

\bibitem{Monchaux08_prl} R. Monchaux, P.-P. Cortet, P.-H. Chavanis, A. Chiffaudel, F. Daviaud, P. Diribarne, B. Dubrulle, Phys. Rev. Lett. \textbf{101}, 174502 (2008)

\bibitem{Renouf05_pof} M. Renouf, D. Bonamy, F. Dubois, P. Alart, Phys. Fluids \textbf{17}(10), 103303 (2005)

\bibitem{Rajchenbach00_ap} J. Rajchenbach, Adv. Phys. \textbf{49}, 229 (2000)

\bibitem{Bonamy02_pof} D. Bonamy, F. Daviaud, L. Laurent, Phys. Fluids \textbf{14}(5), 1666 (2002)

\bibitem{Bonamy03_gm} D. Bonamy, F. Daviaud, L. Laurent, P. Mills, Gran. Matt. \textbf{4}, 183 (2003)

\bibitem{Jop05_jfm} P. Jop, Y. Forterre, O. Pouliquen, J. Fluid Mech. \textbf{541}, 167 (1990)

\bibitem{Speedy99_jcp} R.J. Speedy, J. Chem. Phys. \textbf{110}, 4559 (1999)

\bibitem{Chavanis97_prl} P.-H. Chavanis,  J. Sommeria, Phys Rev. Lett. \textbf{78}, 3302 (1997)

\bibitem{Chavanis02_pre} P.-H. Chavanis,  J. Sommeria, Phys Rev. E \textbf{65}, 026302 (2002)

\bibitem{Moreau88_proc} J.-J. Moreau, in \textit{Non Smooth Mechanics and Applications, CISM Courses and
Lectures}, edited by P.-D. Panagiotopoulos (Springer-Verlag, Wien,
New York, 1988), p. 1

\bibitem{Jean99_cmame} M. Jean, Comp. Meth. Appl. Mech. Engrg. \textbf{177}, 235 (1999)

\bibitem{Renouf04_cmame} M. Renouf, P. Alart, Comp. Meth. Appl. Mech. Engrg. \textbf{194}, 2019 (2004)

\bibitem{Renouf04_jcam} M. Renouf, F. Dubois, P. Alart, J. Comput. Appl. Math. \textbf{168}, 375 (2004)

\bibitem{Bonamy02_prl} D. Bonamy, F. Daviaud, L. Laurent, M. Bonetti, J.-P. Bouchaud, Phys. Rev. Lett. \textbf{89}, 034301 (2002)

\bibitem{Reynolds85_pms} O. Reynolds, Phyl. Mag. Ser. 5 \textbf{20}, 469 (1885)

\bibitem{Komatsu01_prl} T.S. Komatsu, S. Inagasaki, N. Nakagawa, S. Nasuno, Phys. Rev. Lett. \textbf{86}, 1757 (2001)

\bibitem{CourrechdePont05_prl} S. Courrech du Pont, R. Fisher, P. Gondret, B. Perrin, M. Rabaud, Phys. Rev. Lett. \textbf{94}, 048003 (2005)

\bibitem{Crassous} J. Crassous, J.-F. Metayer, P. Richard, C. Laroche, J. Stat. Mech., P03009 (2008).

\end{thebibliography}
\end{document}